\documentclass[journal=jacsat,manuscript=article]{achemso}

\usepackage[version=3]{mhchem} % Formula subscripts using \ce{}

\author{Oğuzhan Yücel}
\email{oguzhan.yuecel@fu-berlin.de}
\affiliation[1]{Department of Physics, Freie Universit\"at Berlin, 14195 Berlin, Germany}
\author{Denis Yagodkin}
\affiliation[1]{Department of Physics, Freie Universit\"at Berlin, 14195 Berlin, Germany}
\author{Jan N. Kirchhof}
\affiliation[1]{Department of Physics, Freie Universit\"at Berlin, 14195 Berlin, Germany}
\author{Yuefeng Yu}
\affiliation[1]{Department of Physics, Freie Universit\"at Berlin, 14195 Berlin, Germany}
\author{Abhijeet M. Kumar}
\affiliation[1]{Department of Physics, Freie Universit\"at Berlin, 14195 Berlin, Germany}
\author{Adrian Dewambrechies}
\affiliation[1]{Department of Physics, Freie Universit\"at Berlin, 14195 Berlin, Germany}
\author{Sviatoslav Kovalchuk}
\affiliation[1]{Department of Physics, Freie Universit\"at Berlin, 14195 Berlin, Germany}
\author{Kirill I. Bolotin}
\email{kirill.bolotin@fu-berlin.de}
\affiliation[1]{Department of Physics, Freie Universit\"at Berlin, 14195 Berlin, Germany}

\title[An \textsf{achemso} demo]
  {Strain activation of localized states in {\ce{WSe2} }}

\begin{document}

\begin{abstract}
	
Single-photon emission centers generated by controlled atomic force microscopy (AFM) indentation in monolayer \ce{WSe2} on a flexible polymer substrate are explored for applications in quantum technologies. Here, we study the response of these emitters to the polymer substrate's strain state, which is controlled by selecting the indentation force and by gradually thermally annealing the samples. In the indented areas, we observe sharp new photoluminescence (PL) peaks in the regions 1.62 $-$ 1.68 eV and 1.70 $-$ 1.73 eV characterized by sublinear power dependence and spectral wandering. We find that these peaks arise only when the indentation force exceeds a few $\mu$N and generally red-shift as the applied force increases. Conversely, after thermal annealing (T $< 60$ $^{\circ}$C), \ce{WSe2} experiences strain relaxation, leading to a blue shift of the peaks' spectral position and their ultimate disappearance. Our analysis of the peaks’ positions vs. strain allows us to draw several conclusions about the nature of these emission. Specifically, we elucidate the roles of excitonic confinement and hybridization between free excitons and defect-related states, a process activated by the strain level. Overall, our approach suggests that the energy of localized emitters may be controlled via strain engineering.

\end{abstract}

\section{Introduction}

Two-dimensional semiconductors from the group of transition metal dichalcogenides (TMDs) support a diverse zoo of excitons with varying properties \cite{wang2018colloquium, regan2022, mueller2018, berkelbach2013}. Earlier studies of the excitonic landscape focused on free excitons, such as neutral excitons, charged excitons, biexcitons, and other species, often considering disorder-localized excitonic states as artifacts of material imperfections\cite{Noll1990, Webb1991}. More recently, localized excitonic states distinguished by their sublinear power dependence became a budding research topic in their own right \cite{von1990localized}. First, localized excitonic states have been found to “store” spin and valley information in TMDs\cite{Linhart2019, wang2020spin}. Second, localized states have been employed to create single photon emitters (SPEs) in TMDs\cite{palacios2018atomically,abramov2023}. While the coherence or stability properties of such emitters are inferior to those of more established technologies, e.g. NV centers in diamond\cite{jelezko2001}, TMD SPEs are distinguished by their spatial position directly at the material's surface, enabling coupling to external objects (e.g., for sensing or enhancement), electrical tunability stemming from the semiconductor nature of the TMDs, and the variability of emission energy\cite{wang2018electrical, zheng2018light}.

In general, it is well-established that SPEs appear in several materials from the TMD family, notably in \ce{WSe2}, under the application of inhomogeneous strain\cite{palacios2017,branny2017}. Such strain fields can be applied by depositing TMDs onto pre-patterned substrates\cite{branny2017, palacios2017} or by indenting TMDs with sharp atomic force microscope (AFM) tips\cite{abramov2023,rosenberger2019quantum, so2021electrically}. The strain-induced SPEs are heralded by the appearance of sharp new peaks in photoluminescence spectra characterized by narrow spectral width and sublinear power dependence. Electrical control, energy tunability, and coupling to dielectric and plasmonic antennas have all been demonstrated for TMD SPEs\cite{Azzam2023, Patel2023, Aharonovich2016, palacios2018atomically, clark2016, Peng2020}.

Despite the recent surge of interest in strain-related localized SPEs in TMDs and the ubiquity of approaches to produce them, their physical origins remain a topic of debate\cite{abramov2023}. First, while some studies suggest that excitons are localized by strain-induced potential, others suggest that otherwise "dark" free excitons radiatively brightened in the presence of defects activated with strain are necessary for the formation of emitters\cite{tongay2013defects, Srivastava2015, chen2023activated}. It has also been suggested that local straining techniques can produce extrinsic structural defects\cite{Parto2021, palacios2017, branny2017}. Second, the role of strain-related exciton “funneling” in brightening the emitters remains debated\cite{hernandez2022strain, vutukuru2021enhanced, harats2020}. Finally, it is still unclear whether the energy of the emitters can be controlled in a large spectral range, e.g. via variation of the strain magnitude\cite{Iff2019}. 
 
Our work is devoted to answering some of these questions. We study strain-induced localized emitters by controllably indenting \ce{WSe2} monolayers transfered onto a polymer substrate. We show that AFM indentation process pierces monolayer TMDs during the process of strain transfer and determine an optimal range of indentation forces to observe the emitters. We then demonstrate that the strain induced in a specific emitter can be relaxed through thermal annealing at moderate temperature levels or completely removed at higher temperatures approaching the glass transition temperature of poly methyl-methacrylate polymer (PMMA), (Fig. S3). Finally, we analyze the spectral shifts and intensity enhancements of the emitters. Our findings are consistent with the emission originating from the hybridization between free excitons and defect states. Furthermore, our results highlight a potential pathway for achieving spectrally tunable single-photon emitters (SPEs).

\textbf{Experimental setup and measurement:} We induce inhomogeneous strain by indenting a TMD on an elastic substrate using an AFM tip, following the approach of Ref.\cite{rosenberger2019quantum, so2021electrically}. The substrate consists of Si/SiO$_2$ substrate coated with 300 nm of PMMA. We then mechanically exfoliate monolayer \ce{WSe2} onto these substrates. We indent these samples by applying controlled forces (Fig. 2, S4) and record changes in photoluminescence (PL) spectra from the same sample placed in an optical cryostat before and after indentation at $T= 4$ K. We avoid regions of the flake with irregularities, such as wrinkles or cracks (Fig. S2). We use an excitation laser with 532 nm wavelength at varying power $P=10$ nW to 10 $\mu$W to resolve localized features. Before the indentation (Fig. 1c, black), we observe excitonic features routinely seen in \ce{WSe2} monolayers $-$ a neutral exciton (X$_0$) at 1.75 eV and features typically associated with other excitonic states and disorder below 1.7 eV \cite{wang2018colloquium, kirchhof2022nanomechanical}. We observe the appearance of new features after the indentation with a maximum force of $5.5$ $\mu$N, in more than half of all measured indented regions. These peaks generally emerge at device-specific energies below 1.75 eV. In a representative sample, for instance, we observe two new features at 1.718 eV and 1.670 eV (Fig. 1c, red).

These new peaks have several key characteristic traits \cite{warburton2000optical, tran2016quantum}. First, the peaks are sharp, e.g., full width at half maximum (FWHM) of $1.3$ meV for the peak at 1.718 eV in Fig. 1c, smaller than that for other free and disorder-related excitons (e.g., FWHM=13 meV for $X_0$). Second, the PL intensity ($I_{PL}$) exhibits sublinear power $(P)$ dependence, $I_{PL} \approx$ {$P^\alpha$}, with \textit{$\alpha$} between 0.5 – 0.7, consistent with previous studies of indented samples \cite{Srivastava2015, He2016} (Fig. 1d). Third, the peak is only visible at low temperatures and disappears when $ T >30$ K. Finally, the emission arises in a spatially localized region with the size below the diffraction limit (Fig. S1). These features have been previously observed in locally strained samples by multiple groups and explained as strain-induced localized states exhibiting single-photon emitter behavior \cite{Iff2019, darlington2020, kim2019, aslan2018strain}. We therefore conclude that our AFM indentation approach also generates strain-related localized states.

\textbf{Exploring localized emitters:} Next, we explore the effect of the indentation conditions on the properties of the SPEs. An observable indentation appears at around 1 $\mu$N indentation force. The depth of the indentation increases with maximum applied force (Fig. 2b). It is noteworthy that in all devices with detectable indentation, we observe a sharp jump in the force-displacement curves during the indentation at around 0.5 $\mu$N (Fig. 2a). This jump corresponds to the AFM tip puncturing the \ce{WSe2}  flake. Indeed, we find that the slope of the force-indentation curve -- which reflects the stiffness of the indented material -- is higher than the value corresponding to that of bare PMMA (black curve in Fig. 2a). This mismatch corresponds to the contribution from the stiffness of the \ce{WSe2}. At the critical force value, the slope returns to the PMMA value (inset in Fig. 2a); the slope remains roughly constant during AFM tip withdrawal. We also note the deviation from linearity in the indentation curve above 3 $\mu$N. This deviation signals the formation of a pile-up of PMMA at the corners of the indented region.

We then examine the interrelation between the indentation force and SPE features (Fig. 3). We find that new indentation-related peaks arise in three spectral regions marked yellow in Fig. 3a. In general, peaks matching characteristics of SPEs (e.g., exhibiting spectral wandering, Fig. 3c) appear for indentation force higher than 1 $\mu$N. We observe SPEs in low-energy regions,  1.63–1.66 eV and 1.54–1.58 eV, for indentation forces higher than 3.5 $\mu$N. While the observed SPEs are stable at low temperatures or between subsequent cooldowns, they change drastically under mild thermal annealing, e.g., 60 $^{\circ}$C in Fig. 3b. A general trend that we note is that the SPEs reduce in intensity and disappear entirely from the lowest spectral region. The results of Figs. 2 and 3 are consistent with mechanical strain -- and not just the structural defects in \ce{WSe2}  -- playing a critical role in the formation of SPEs. The appearance of SPEs only for large indentation forces is consistent with the large local strain being needed for their formation. Conversely, the disappearance of SPEs after annealing is consistent with thermally-assisted strain relaxation.

To investigate this strain relaxation more closely, we study three regions indented using a force of 5.5 $\mu$N.  While the emission from the indented region is stable over days, we noticed significant changes over longer periods. After two months of storage under ambient conditions, the indentation-related peaks initially observed around 1.645 eV disappeared from this spectral range (Fig. 4b) while in some samples new peaks appeared at higher energies (1.715 eV). The plot colors (Fig. 4b) indicate the annealing history of the sample that defines its strain state (red is right after indentation, orange is the sample stored at max. 30 $^\circ$C for two months, and blue is the sample annealed at 120 $^\circ$C). All three spectral measurements are recorded at 4 Kelvin after performing the thermal treatment. These changes are accompanied by a reduction in the indentation depth, e.g., from 33 nm to 27 nm, Fig. 4a. This behavior strongly suggests that the glassy flow of the PMMA polymer leads to strain relaxation at the \ce{WSe2}/PMMA interface, which directly modifies the properties of localized excitonic potentials. Over time, strain redistribution diminishes the confinement of excitons, resulting in both a spectral shift and the eventual disappearance of the strain-induced emission features. Such a flow is dramatically accelerated at elevated temperatures. Indeed, after a subsequent brief (5 minutes) heating to 120 $^\circ$C, the indentation depth of the same device in Fig. 4a reduced from 27 nm to 19 nm while the indentation-related peaks disappeared almost entirely in all three emission centers (Fig. 4b). The accelerated strain relaxation at higher temperatures reflects the thermally activated flow of the PMMA, allowing the lattice to recover to a near-unstrained state. Consequently, the localized strain potentials are eliminated, enabling trapped excitons to delocalize. 

To quantify the effect of thermally induced changes on SPEs, we analyze the statistical changes in the PL spectra before and after annealing (Fig. 5). To visualize indentation-related peaks in many regions, we chose to represent the PL peaks by their amplitudes $\eta$, evaluated as the area under the corresponding peak (obtained from a Gaussian fit to the data) normalized by its FWHM. To exclude free-excitonic features, we only analyze the peaks with FWHM twice smaller than that for free excitons. To facilitate the comparison of various peaks across different spectra, we perform a spectral shift and alignment procedure, ensuring that all spectra were shifted to ensure that $X_0$ is positioned at 1.750 eV. In general, peaks with large $\eta$ correspond to intense and narrow features. In Fig. 3, we plotted the amplitudes $\eta$ of all peaks in the 25 spots before and after indentation, as well as after annealing. These results confirm the behavior already seen in the representative device in Fig. 4b. Before indentation, we see the majority of excitonic peaks in the region 1.668 – 1.702 eV (white region). These peaks not caused by indentation are likely related to preexisting disorders and/or localized strain, therefore, are not analyzed further.  Directly after the indentation, we predominantly observe new intense peaks in the region $<$1.668 eV and $>$1.7 (yellow-shaded region). After the annealing, the number of peaks in the region $<$1.668 eV sharply drops, whereas some sharp features emerge in the region $>$1.702 eV.

{\bf Discussion:} The observations of Figs. 2–5 allow us to draw several conclusions regarding the nature of strain-induced localized peaks. First,  we show that the indentation required to generate SPEs punctures \ce{WSe2}. Nevertheless, structural defects potentially produced by such puncturing are not the exclusive reason behind SPEs appearance. Indeed, if that were the case, we expect the SPE to appear even for small indentation forces and persist after the strain in the flake is relaxed after thermal annealing. Both of these expectations contradict the observations in Figs. 2–5. Second, the character of localized emitters is inconsistent with a commonly assumed model -- free excitons confined by the localized potential near the strain maximum in the indented region. If that were the case, the peaks due to localized emitters would emerge at spectral positions entirely defined by the localized strain level. We would then observe such peaks throughout a broad spectral range defined by the maximum strain. Instead, we observe the SPE peaks predominantly inside narrow spectral regions (yellow bands in Fig. 5). In contrast, our observations are consistent with normally dark structural defects brightening under strain through hybridization with free excitons \cite{hernandez2022strain, moon2020strain, abramov2023, Linhart2019}. We note while compressive strains can result in a direct-to-indirect bandgap transition, the tensile strain applied in our setup preserves the direct nature of the bandgap\cite{nr2015,nl2013,npj2D2017,aip2022, rsc2014}. Our data suggest that there are several different types of structural defects, which can all couple to free excitons depending on the amount of strain. Indeed, when the strain is large, we see sharp peaks predominantly in the region 1.63 – 1.668 eV, and in the region 1.702 – 1.732 eV for lower strain, hinting that different emitter types are 'responsible' for these regions. Finally, we propose that strain-related funneling is at least partially responsible for the high brightness of the localized emitters. Indeed, we observe brighter emission from the emitters in high-strain devices (red points in Fig. 5). We note that other potential mechanisms, including position-dependent doping can contribute to emitters' brightness\cite{harats2020}.

To summarize, we investigated statistical changes in strain-induced localized emitters in indented monolayer \ce{WSe2}. We found that the average strain state of emitters can be tuned by heating the PMMA substrate due to the nature of PMMA substrate. By analyzing changes in the optical spectra of individual emitters vs. indentation force and annealing conditions, we can draw several conclusions regarding the emitter’s character. In particular, our data is consistent with several different types of structural emitters being hybridized with free excitons depending on the strain state. Finally, our approach suggests that the energy of localized states may be controlled via strain engineering. This capability may prove useful in the emerging field of quantum technologies. At the same time, gradual strain relaxation in PMMA-based samples suggests that different types of elastic materials may be needed to enable functional indentation-based devices. 

\begin{acknowledgement}

The authors thank Bianca Höfer for facility management. We acknowledge financial support through BMBF, German Federal Ministry for Education and Research, project 05K22KE3, and DFG, German Research Foundation, Project ID: 449596295 and TRR227 B08, 328545488, and SPP2244. We thank Pablo Lopez and Dr. Sebastian Heeg for the experimental help.

\end{acknowledgement}

\newpage
%%%%%%%%%%%%%%%%%%%%%%%%%%%%%%%%%%%%%%%%%%%%%%%%%%%%%%%%%%%%%%%%%%%%%
%%%%%%%%%%%%%%%%%%%%%%%%%%% MAIN FIGURES
\begin{figure}[ht]
	\centering
	\includegraphics[width=1\textwidth]{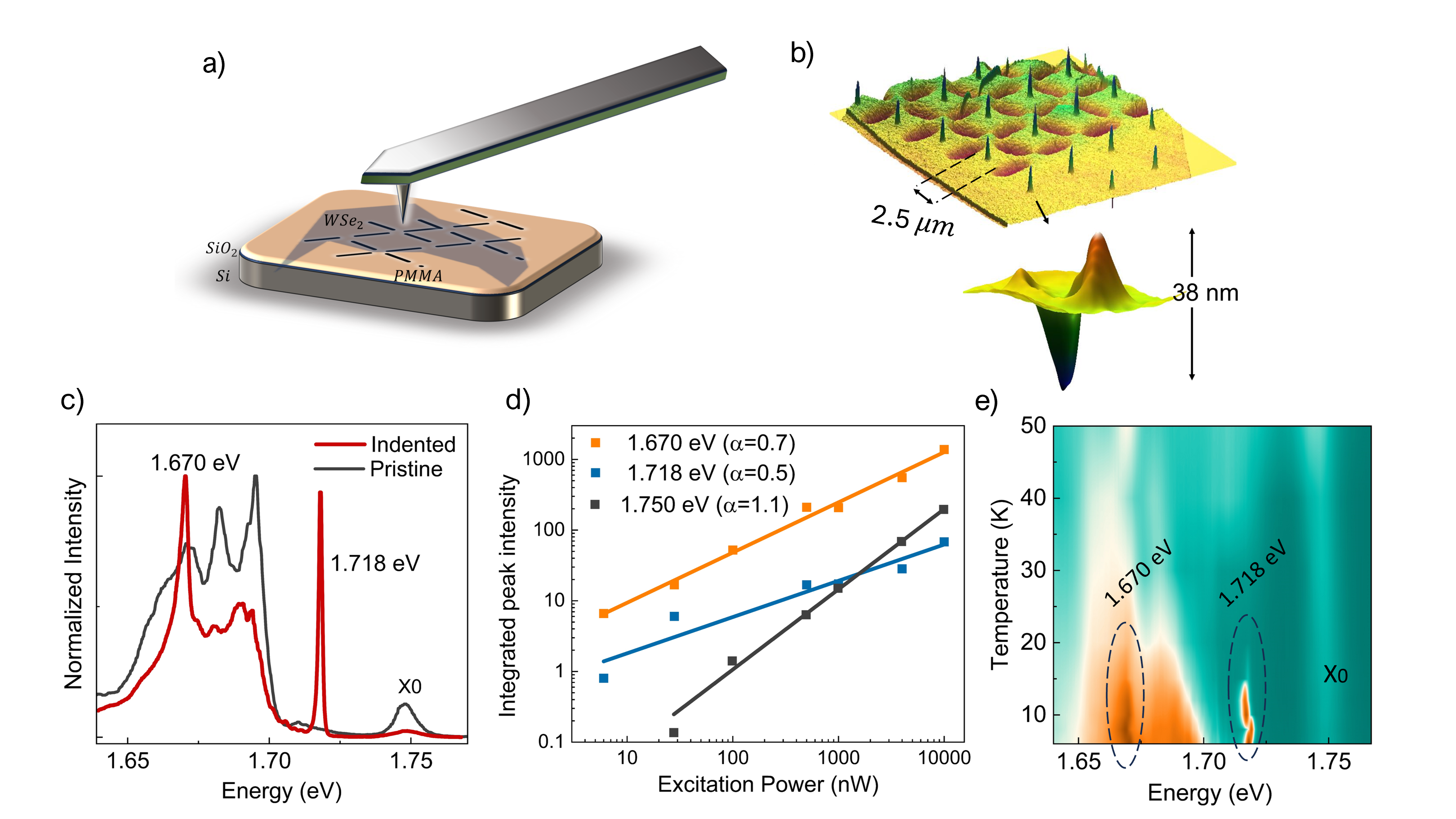}
	\caption{\textbf{Creation of position-controllable localized emission centers in\\\ce{WSe2}/PMMA} a) Cartoon of the setup. A \ce{WSe2}/PMMA sample is locally indented by an AFM tip, b) AFM images of the indented sample used for the extraction of the indentation depth. c) Photoluminescence spectra at T=6 K from the same sample point before (black) and after (red) indentation. New sharp PL peaks 1.670 eV and 1.718 eV appearing after indentation are prominent. d), e) Power and temperature dependencies of the peaks at 1.670 eV and 1.718 eV, respectively, are consistent with their localized character.}
\end{figure}
%%%%%%%%%%%%%%%%%%%%%%%%%%%%%%%%%%%%%%%%%%%%%%%
\newpage
\begin{figure}[ht]
	\centering
	\includegraphics[width=1\textwidth]{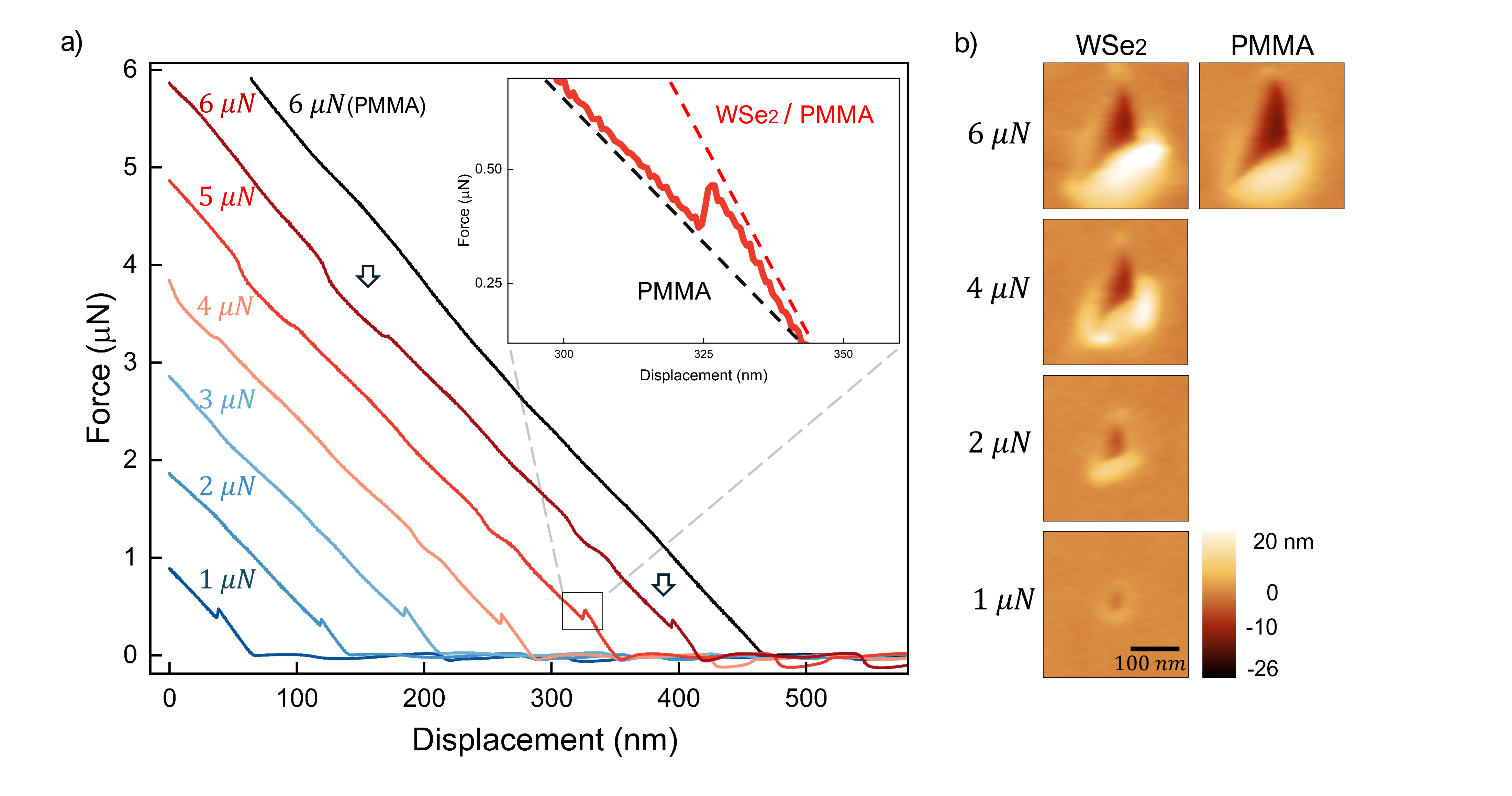}
	\caption{\textbf{Comparative indentation force analysis} a) The force-displacement curves for \ce{WSe2} monolayer on PMMA, and bare PMMA substrate (black) ranging from 1 $\mu$N to 6 $\mu$N. Inset: The material’s breakage range as zoom-in view for 5 $\mu$N. Dashed black line indicates the slope of PMMA reference line and the red dots indicate the slope of \ce{WSe2} line. b) AFM images for indentation made at incremental forces. Morphological changes induced by increasing applied force provide insights into the mechanical response and deformation characteristics of the materials.}
\end{figure}

\newpage
\begin{figure}[ht]
	\centering
	\includegraphics[width=1\textwidth]{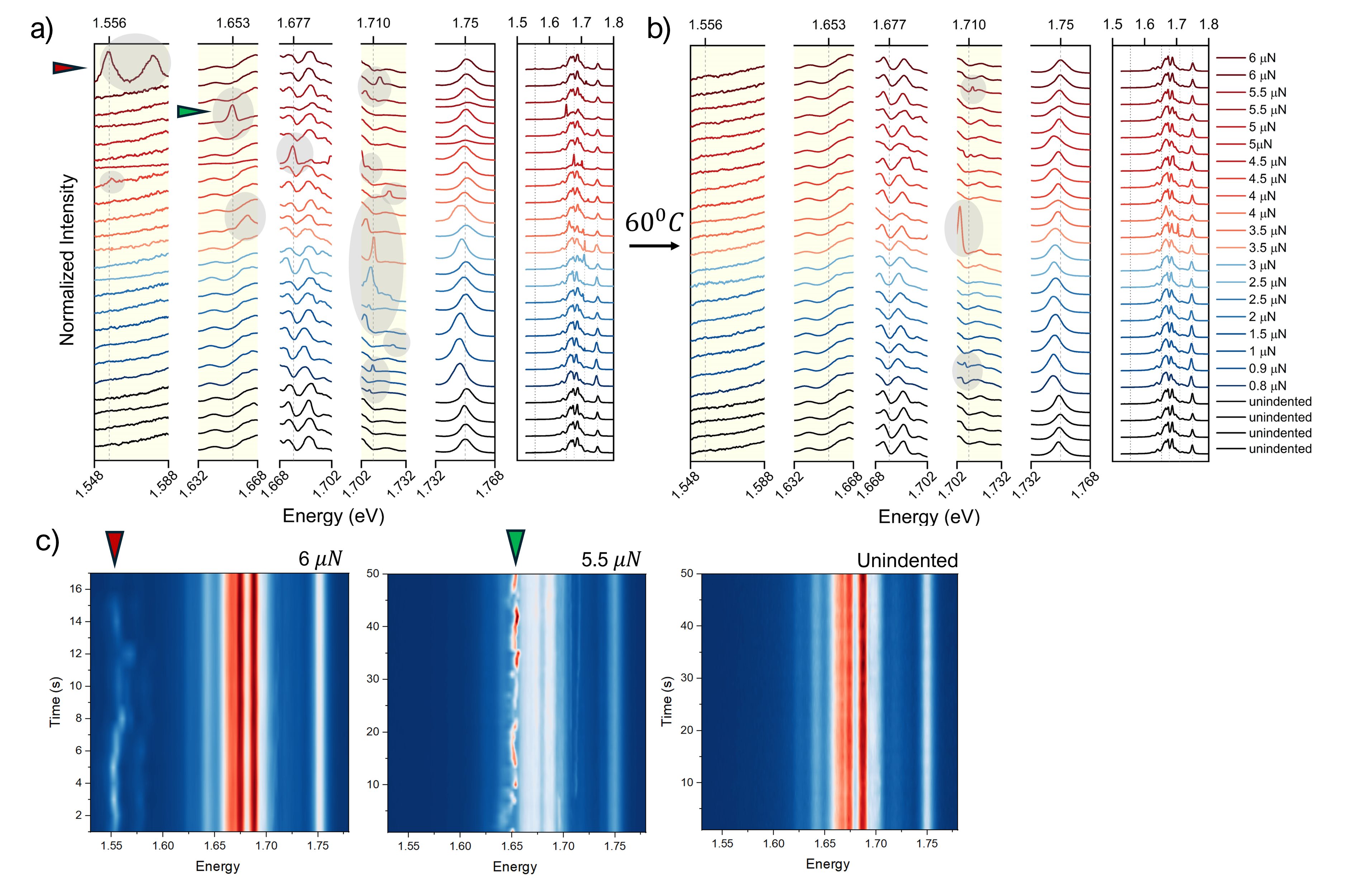}
	\caption{ \textbf{\textbf{Photoluminescence (PL) spectra of varying indentation forces before and after annealing}. } a) The initial PL measurements at 4 K for incremental indentation forces ranging from 0.8 $\mu$N to 6 $\mu$N. Distinct energy shifts and intensity variations are highlighted across specific energy ranges with notable changes marked in shaded regions. b) The PL spectra measured at 4 K after annealing the sample at 335 K (60 $^{\circ}$C), illustrating how thermal treatment influences the relaxation of strain, altering peak positions and intensities. Full-range PL spectra are included for completeness in the right side of each panel. c) The time-dependent PL spectra for several representative devices see in (a).}
\end{figure}

%%%%%%%%%%%%%%%%%%%%%%%%%%%%%%%%%%%%%%%%%%%%%%
\newpage
\begin{figure}[ht]
	\centering
	\includegraphics[width=1\textwidth]{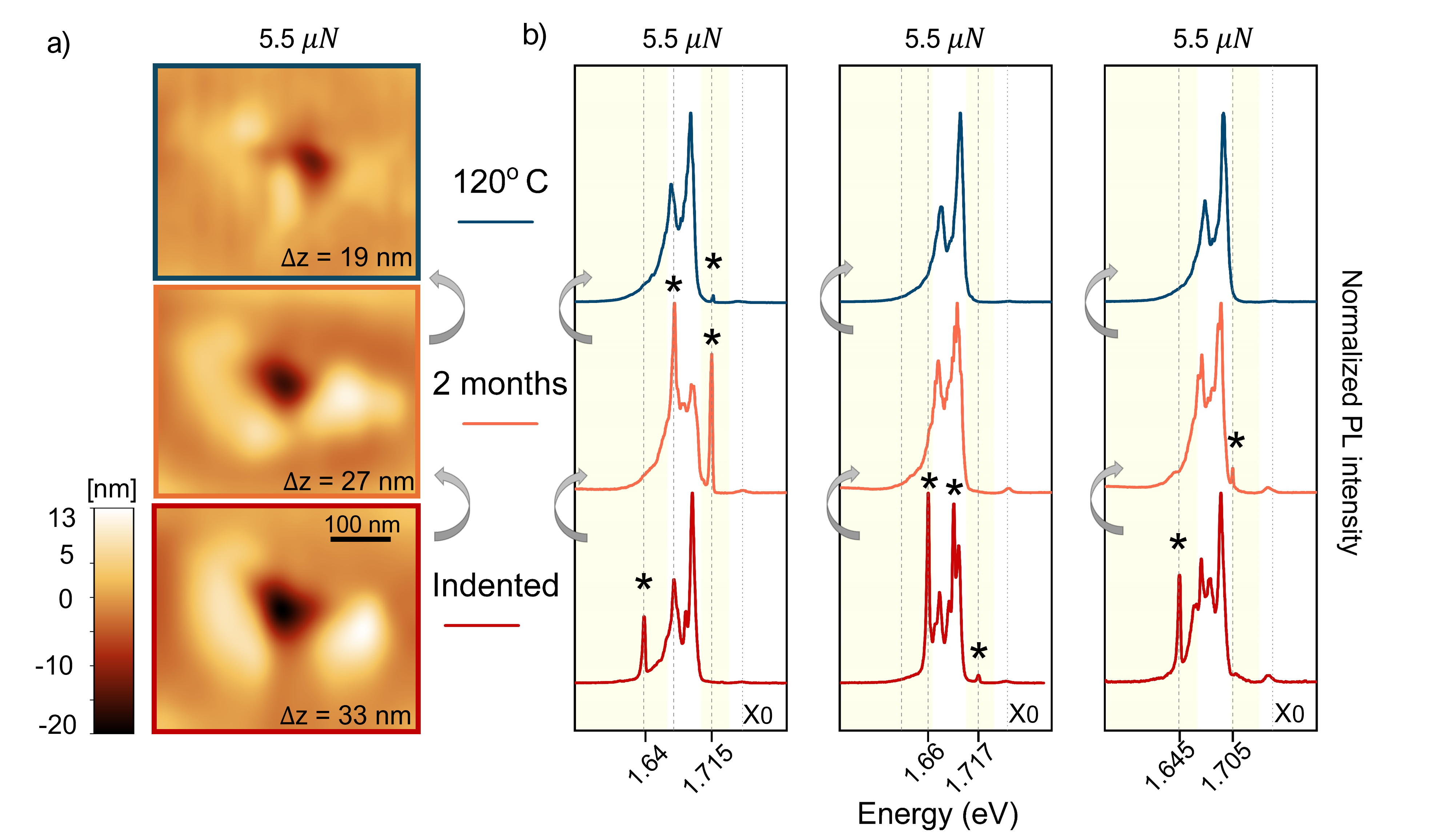}
	\caption{ \textbf{Controlling the strain state of the sample via annealing}. AFM images at the same region (a) and PL spectra of three representative emitters (b) after indentation (red), two months later at ambient condition (orange), and after subsequent 5 min annealing at 120 $^{\circ}$C (blue). Strain relaxation after each step is apparent in a). Indentation-related peaks appearing inside the shaded regions and blue-shifting at each stage of the process are seen in b).}
\end{figure}

%%%%%%%%%%%%%%%%%%%%%%%%%%%%%%%%%%%%%%%%%%%%%%
\newpage
\begin{figure}[ht]
	\centering
	\includegraphics[width=1\textwidth]{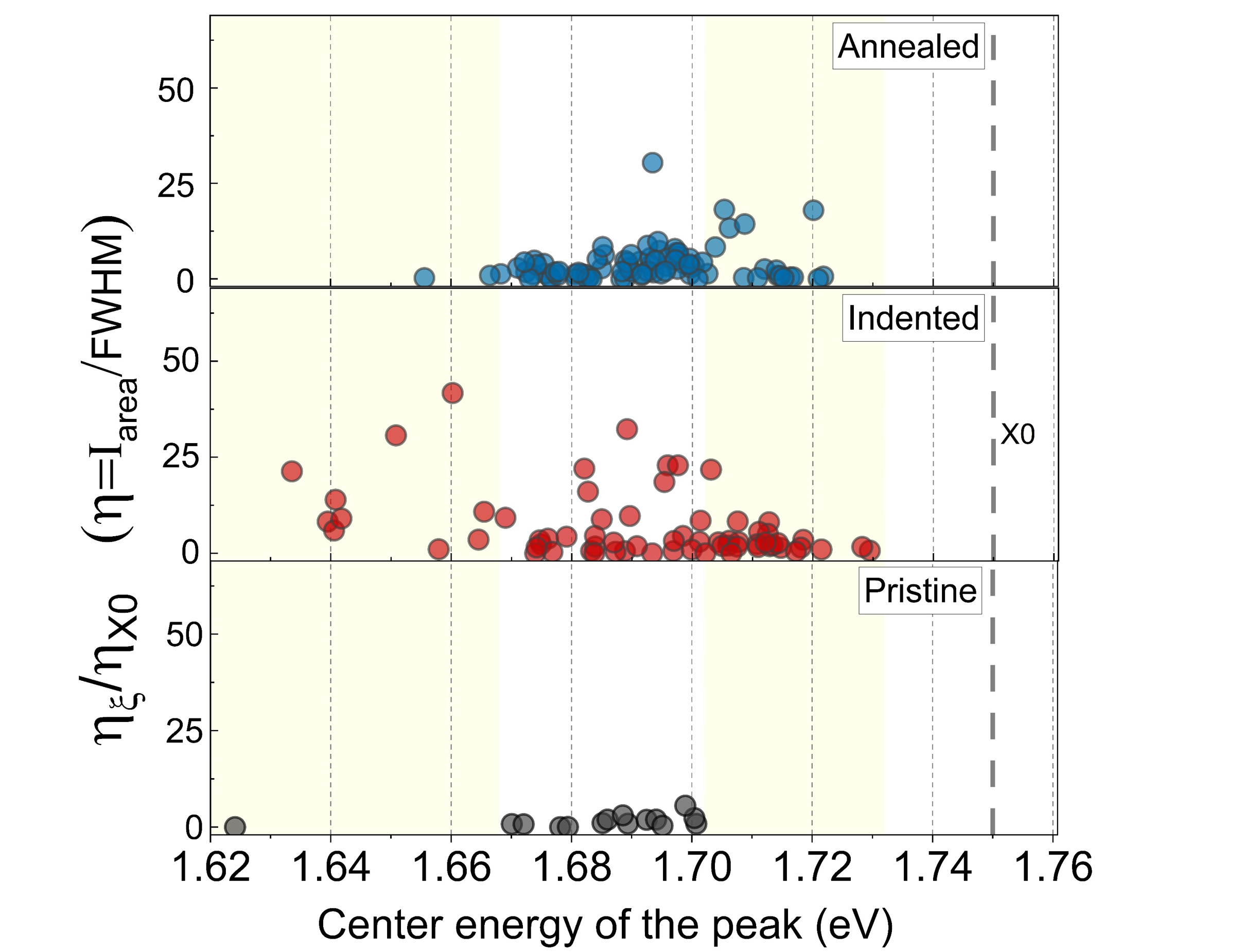}
	\caption{ \textbf{Evolution of PL during indentation and relaxation.} The peaks “heights” (defined as area under the peak/FWHM) vs. peak central energy for 25 indentation spots before the indentation, after the indentation, as well as after temperature cycling. The maximum indentation force is $5 - 5.5 \mu$N in all cases. To compare the spectra from different devices and account for substrate-wide doping and strain variations, each spectrum is aligned according to X$_0$ =1.750 eV and normalized by their own intensity of corresponding X$_0$ peak. }
\end{figure}

\newpage
\bibliography{ref}

\providecommand{\latin}[1]{#1}
\makeatletter
\providecommand{\doi}
  {\begingroup\let\do\@makeother\dospecials
  \catcode`\{=1 \catcode`\}=2 \doi@aux}
\providecommand{\doi@aux}[1]{\endgroup\texttt{#1}}
\makeatother
\providecommand*\mcitethebibliography{\thebibliography}
\csname @ifundefined\endcsname{endmcitethebibliography}
  {\let\endmcitethebibliography\endthebibliography}{}
\begin{mcitethebibliography}{45}
\providecommand*\natexlab[1]{#1}
\providecommand*\mciteSetBstSublistMode[1]{}
\providecommand*\mciteSetBstMaxWidthForm[2]{}
\providecommand*\mciteBstWouldAddEndPuncttrue
  {\def\EndOfBibitem{\unskip.}}
\providecommand*\mciteBstWouldAddEndPunctfalse
  {\let\EndOfBibitem\relax}
\providecommand*\mciteSetBstMidEndSepPunct[3]{}
\providecommand*\mciteSetBstSublistLabelBeginEnd[3]{}
\providecommand*\EndOfBibitem{}
\mciteSetBstSublistMode{f}
\mciteSetBstMaxWidthForm{subitem}{(\alph{mcitesubitemcount})}
\mciteSetBstSublistLabelBeginEnd
  {\mcitemaxwidthsubitemform\space}
  {\relax}
  {\relax}

\bibitem[Wang \latin{et~al.}(2018)Wang, Chernikov, Glazov, Heinz, Marie, Amand,
  and Urbaszek]{wang2018colloquium}
Wang,~G.; Chernikov,~A.; Glazov,~M.~M.; Heinz,~T.~F.; Marie,~X.; Amand,~T.;
  Urbaszek,~B. Colloquium: Excitons in atomically thin transition metal
  dichalcogenides. \emph{Reviews of Modern Physics} \textbf{2018}, \emph{90},
  021001\relax
\mciteBstWouldAddEndPuncttrue
\mciteSetBstMidEndSepPunct{\mcitedefaultmidpunct}
{\mcitedefaultendpunct}{\mcitedefaultseppunct}\relax
\EndOfBibitem
\bibitem[Regan \latin{et~al.}(2022)Regan, Wang, Paik, Zeng, Zhang, Zhu,
  MacDonald, Deng, and Wang]{regan2022}
Regan,~E.~C.; Wang,~D.; Paik,~E.~Y.; Zeng,~Y.; Zhang,~L.; Zhu,~J.;
  MacDonald,~A.~H.; Deng,~H.; Wang,~F. Emerging exciton physics in transition
  metal dichalcogenide heterobilayers. \emph{Nature Reviews Materials}
  \textbf{2022}, \emph{7}, 778--795\relax
\mciteBstWouldAddEndPuncttrue
\mciteSetBstMidEndSepPunct{\mcitedefaultmidpunct}
{\mcitedefaultendpunct}{\mcitedefaultseppunct}\relax
\EndOfBibitem
\bibitem[Mueller and Malic(2018)Mueller, and Malic]{mueller2018}
Mueller,~T.; Malic,~E. Exciton physics and device application of
  two-dimensional transition metal dichalcogenide semiconductors. \emph{npj 2D
  Materials and Applications} \textbf{2018}, \emph{2}, 29\relax
\mciteBstWouldAddEndPuncttrue
\mciteSetBstMidEndSepPunct{\mcitedefaultmidpunct}
{\mcitedefaultendpunct}{\mcitedefaultseppunct}\relax
\EndOfBibitem
\bibitem[Berkelbach \latin{et~al.}(2013)Berkelbach, Hybertsen, and
  Reichman]{berkelbach2013}
Berkelbach,~T.~C.; Hybertsen,~M.~S.; Reichman,~D.~R. Theory of neutral and
  charged excitons in monolayer transition metal dichalcogenides.
  \emph{Physical Review B} \textbf{2013}, \emph{88}, 045318\relax
\mciteBstWouldAddEndPuncttrue
\mciteSetBstMidEndSepPunct{\mcitedefaultmidpunct}
{\mcitedefaultendpunct}{\mcitedefaultseppunct}\relax
\EndOfBibitem
\bibitem[Noll \latin{et~al.}(1990)Noll, Ploog, Brunner, and Feenstra]{Noll1990}
Noll,~G.; Ploog,~K.; Brunner,~K.; Feenstra,~R.~M. Picosecond Stimulated Photon
  Echo Due to Intrinsic Excitations in Semiconductor Mixed Crystals.
  \emph{Physical Review Letters} \textbf{1990}, \emph{64}, 792--795\relax
\mciteBstWouldAddEndPuncttrue
\mciteSetBstMidEndSepPunct{\mcitedefaultmidpunct}
{\mcitedefaultendpunct}{\mcitedefaultseppunct}\relax
\EndOfBibitem
\bibitem[Webb \latin{et~al.}(1991)Webb, Shah, Sugahara, Sivco, and
  Cho]{Webb1991}
Webb,~M.~D.; Shah,~J.; Sugahara,~H.; Sivco,~D.~L.; Cho,~A.~Y. Observation of
  Time-Resolved Picosecond Stimulated Photon Echoes and Free Polarization Decay
  in GaAs/AlGaAs Multiple Quantum Wells. \emph{Physical Review Letters}
  \textbf{1991}, \emph{66}, 934--937\relax
\mciteBstWouldAddEndPuncttrue
\mciteSetBstMidEndSepPunct{\mcitedefaultmidpunct}
{\mcitedefaultendpunct}{\mcitedefaultseppunct}\relax
\EndOfBibitem
\bibitem[Von~der Osten and Stolz(1990)Von~der Osten, and
  Stolz]{von1990localized}
Von~der Osten,~W.; Stolz,~H. Localized exciton states in silver halides.
  \emph{Journal of Physics and Chemistry of Solids} \textbf{1990}, \emph{51},
  765--791\relax
\mciteBstWouldAddEndPuncttrue
\mciteSetBstMidEndSepPunct{\mcitedefaultmidpunct}
{\mcitedefaultendpunct}{\mcitedefaultseppunct}\relax
\EndOfBibitem
\bibitem[Linhart \latin{et~al.}(2019)Linhart, Paur, Smejkal, Burgd\"orfer,
  Mueller, and Libisch]{Linhart2019}
Linhart,~L.; Paur,~M.; Smejkal,~V.; Burgd\"orfer,~J.; Mueller,~T.; Libisch,~F.
  Localized Intervalley Defect Excitons as Single-Photon Emitters in
  ${\mathrm{WSe}}_{2}$. \emph{Physical Review Letters} \textbf{2019},
  \emph{123}, 146401\relax
\mciteBstWouldAddEndPuncttrue
\mciteSetBstMidEndSepPunct{\mcitedefaultmidpunct}
{\mcitedefaultendpunct}{\mcitedefaultseppunct}\relax
\EndOfBibitem
\bibitem[Wang \latin{et~al.}(2020)Wang, Deng, Wei, Wan, Liu, Lu, Li, Bi, Zhang,
  Lu, \latin{et~al.} others]{wang2020spin}
Wang,~Y.; Deng,~L.; Wei,~Q.; Wan,~Y.; Liu,~Z.; Lu,~X.; Li,~Y.; Bi,~L.;
  Zhang,~L.; Lu,~H.; others Spin-valley locking effect in defect states of
  monolayer MoS2. \emph{Nano letters} \textbf{2020}, \emph{20},
  2129--2136\relax
\mciteBstWouldAddEndPuncttrue
\mciteSetBstMidEndSepPunct{\mcitedefaultmidpunct}
{\mcitedefaultendpunct}{\mcitedefaultseppunct}\relax
\EndOfBibitem
\bibitem[Palacios-Berraquero and Palacios-Berraquero(2018)Palacios-Berraquero,
  and Palacios-Berraquero]{palacios2018atomically}
Palacios-Berraquero,~C.; Palacios-Berraquero,~C. Atomically-thin quantum light
  emitting diodes. \emph{Quantum confined excitons in 2-dimensional materials}
  \textbf{2018}, 71--89\relax
\mciteBstWouldAddEndPuncttrue
\mciteSetBstMidEndSepPunct{\mcitedefaultmidpunct}
{\mcitedefaultendpunct}{\mcitedefaultseppunct}\relax
\EndOfBibitem
\bibitem[Abramov \latin{et~al.}(2023)Abramov, Chestnov, Alimova, Ivanova,
  Mukhin, Krizhanovskii, Shelykh, Iorsh, and Kravtsov]{abramov2023}
Abramov,~A.~N.; Chestnov,~I.~Y.; Alimova,~E.~S.; Ivanova,~T.; Mukhin,~I.~S.;
  Krizhanovskii,~D.~N.; Shelykh,~I.~A.; Iorsh,~I.~V.; Kravtsov,~V.
  Photoluminescence imaging of single photon emitters within nanoscale strain
  profiles in monolayer WSe $ \_2$. \emph{arXiv preprint arXiv:2301.09478}
  \textbf{2023}, \relax
\mciteBstWouldAddEndPunctfalse
\mciteSetBstMidEndSepPunct{\mcitedefaultmidpunct}
{}{\mcitedefaultseppunct}\relax
\EndOfBibitem
\bibitem[Jelezko \latin{et~al.}(2001)Jelezko, Tietz, Gruber, Popa, Nizovtsev,
  Kilin, and Wrachtrup]{jelezko2001}
Jelezko,~F.; Tietz,~C.; Gruber,~A.; Popa,~I.; Nizovtsev,~A.; Kilin,~S.;
  Wrachtrup,~J. Spectroscopy of Single N-V Centers in Diamond. \emph{Single
  Molecules} \textbf{2001}, \emph{2}, 255--260\relax
\mciteBstWouldAddEndPuncttrue
\mciteSetBstMidEndSepPunct{\mcitedefaultmidpunct}
{\mcitedefaultendpunct}{\mcitedefaultseppunct}\relax
\EndOfBibitem
\bibitem[Wang \latin{et~al.}(2018)Wang, De~Greve, Jauregui, Sushko, High, Zhou,
  Scuri, Taniguchi, Watanabe, Lukin, \latin{et~al.} others]{wang2018electrical}
Wang,~K.; De~Greve,~K.; Jauregui,~L.~A.; Sushko,~A.; High,~A.; Zhou,~Y.;
  Scuri,~G.; Taniguchi,~T.; Watanabe,~K.; Lukin,~M.~D.; others Electrical
  control of charged carriers and excitons in atomically thin materials.
  \emph{Nature nanotechnology} \textbf{2018}, \emph{13}, 128--132\relax
\mciteBstWouldAddEndPuncttrue
\mciteSetBstMidEndSepPunct{\mcitedefaultmidpunct}
{\mcitedefaultendpunct}{\mcitedefaultseppunct}\relax
\EndOfBibitem
\bibitem[Zheng \latin{et~al.}(2018)Zheng, Jiang, Hu, Li, Zeng, Wang, and
  Pan]{zheng2018light}
Zheng,~W.; Jiang,~Y.; Hu,~X.; Li,~H.; Zeng,~Z.; Wang,~X.; Pan,~A. Light
  emission properties of 2D transition metal dichalcogenides: fundamentals and
  applications. \emph{Advanced Optical Materials} \textbf{2018}, \emph{6},
  1800420\relax
\mciteBstWouldAddEndPuncttrue
\mciteSetBstMidEndSepPunct{\mcitedefaultmidpunct}
{\mcitedefaultendpunct}{\mcitedefaultseppunct}\relax
\EndOfBibitem
\bibitem[Palacios-Berraquero \latin{et~al.}(2017)Palacios-Berraquero, Kara,
  Montblanch, Barbone, Latawiec, Yoon, Ott, Loncar, Ferrari, and
  Atat{\"u}re]{palacios2017}
Palacios-Berraquero,~C.; Kara,~D.~M.; Montblanch,~A. R.-P.; Barbone,~M.;
  Latawiec,~P.; Yoon,~D.; Ott,~A.~K.; Loncar,~M.; Ferrari,~A.~C.;
  Atat{\"u}re,~M. Large-scale quantum-emitter arrays in atomically thin
  semiconductors. \emph{Nature communications} \textbf{2017}, \emph{8},
  15093\relax
\mciteBstWouldAddEndPuncttrue
\mciteSetBstMidEndSepPunct{\mcitedefaultmidpunct}
{\mcitedefaultendpunct}{\mcitedefaultseppunct}\relax
\EndOfBibitem
\bibitem[Branny \latin{et~al.}(2017)Branny, Kumar, Proux, and
  Gerardot]{branny2017}
Branny,~A.; Kumar,~S.; Proux,~R.; Gerardot,~B.~D. Deterministic strain-induced
  arrays of quantum emitters in a two-dimensional semiconductor. \emph{Nature
  communications} \textbf{2017}, \emph{8}, 15053\relax
\mciteBstWouldAddEndPuncttrue
\mciteSetBstMidEndSepPunct{\mcitedefaultmidpunct}
{\mcitedefaultendpunct}{\mcitedefaultseppunct}\relax
\EndOfBibitem
\bibitem[Rosenberger \latin{et~al.}(2019)Rosenberger, Dass, Chuang, Sivaram,
  McCreary, Hendrickson, and Jonker]{rosenberger2019quantum}
Rosenberger,~M.~R.; Dass,~C.~K.; Chuang,~H.-J.; Sivaram,~S.~V.;
  McCreary,~K.~M.; Hendrickson,~J.~R.; Jonker,~B.~T. Quantum calligraphy:
  writing single-photon emitters in a two-dimensional materials platform.
  \emph{ACS nano} \textbf{2019}, \emph{13}, 904--912\relax
\mciteBstWouldAddEndPuncttrue
\mciteSetBstMidEndSepPunct{\mcitedefaultmidpunct}
{\mcitedefaultendpunct}{\mcitedefaultseppunct}\relax
\EndOfBibitem
\bibitem[So \latin{et~al.}(2021)So, Kim, Baek, Jeong, Lee, Huh, Kim, Watanabe,
  Taniguchi, Kim, \latin{et~al.} others]{so2021electrically}
So,~J.-P.; Kim,~H.-R.; Baek,~H.; Jeong,~K.-Y.; Lee,~H.-C.; Huh,~W.; Kim,~Y.~S.;
  Watanabe,~K.; Taniguchi,~T.; Kim,~J.; others Electrically driven
  strain-induced deterministic single-photon emitters in a van der Waals
  heterostructure. \emph{Science Advances} \textbf{2021}, \emph{7},
  eabj3176\relax
\mciteBstWouldAddEndPuncttrue
\mciteSetBstMidEndSepPunct{\mcitedefaultmidpunct}
{\mcitedefaultendpunct}{\mcitedefaultseppunct}\relax
\EndOfBibitem
\bibitem[Azzam \latin{et~al.}(2023)Azzam, Parto, and Moody]{Azzam2023}
Azzam,~S.~I.; Parto,~K.; Moody,~G. Purcell enhancement and polarization control
  of single-photon emitters in monolayer WSe2 using dielectric nanoantennas.
  \emph{Nanophotonics} \textbf{2023}, \emph{12}, 477--484\relax
\mciteBstWouldAddEndPuncttrue
\mciteSetBstMidEndSepPunct{\mcitedefaultmidpunct}
{\mcitedefaultendpunct}{\mcitedefaultseppunct}\relax
\EndOfBibitem
\bibitem[Patel \latin{et~al.}(2023)Patel, Parto, Choquer, Umezawa, Hellman,
  Polishchuk, and Moody]{Patel2023}
Patel,~S.~D.; Parto,~K.; Choquer,~M.; Umezawa,~S.; Hellman,~L.; Polishchuk,~D.;
  Moody,~G. Cavity Optomechanics with WSe2 Single Photon Emitters. CLEO 2023.
  2023; p FW3J.2\relax
\mciteBstWouldAddEndPuncttrue
\mciteSetBstMidEndSepPunct{\mcitedefaultmidpunct}
{\mcitedefaultendpunct}{\mcitedefaultseppunct}\relax
\EndOfBibitem
\bibitem[Aharonovich \latin{et~al.}(2016)Aharonovich, Englund, and
  Toth]{Aharonovich2016}
Aharonovich,~I.; Englund,~D.; Toth,~M. Solid-state single-photon emitters.
  \emph{Nature Photonics} \textbf{2016}, \emph{10}, 631--641\relax
\mciteBstWouldAddEndPuncttrue
\mciteSetBstMidEndSepPunct{\mcitedefaultmidpunct}
{\mcitedefaultendpunct}{\mcitedefaultseppunct}\relax
\EndOfBibitem
\bibitem[Clark \latin{et~al.}(2016)Clark, Schaibley, Ross, Taniguchi, Watanabe,
  Hendrickson, Mou, Yao, and Xu]{clark2016}
Clark,~G.; Schaibley,~J.~R.; Ross,~J.; Taniguchi,~T.; Watanabe,~K.;
  Hendrickson,~J.~R.; Mou,~S.; Yao,~W.; Xu,~X. Single defect light-emitting
  diode in a van der Waals heterostructure. \emph{Nano letters} \textbf{2016},
  \emph{16}, 3944--3948\relax
\mciteBstWouldAddEndPuncttrue
\mciteSetBstMidEndSepPunct{\mcitedefaultmidpunct}
{\mcitedefaultendpunct}{\mcitedefaultseppunct}\relax
\EndOfBibitem
\bibitem[Peng \latin{et~al.}(2020)Peng, Chan, Choo, Odom, Sankaranarayanan, and
  Ma]{Peng2020}
Peng,~L.; Chan,~H.; Choo,~P.; Odom,~T.~W.; Sankaranarayanan,~S. K. R.~S.;
  Ma,~X. Creation of Single-Photon Emitters in WSe2 Monolayers Using
  Nanometer-Sized Gold Tips. \emph{Nano Letters} \textbf{2020}, \emph{20},
  5866--5872\relax
\mciteBstWouldAddEndPuncttrue
\mciteSetBstMidEndSepPunct{\mcitedefaultmidpunct}
{\mcitedefaultendpunct}{\mcitedefaultseppunct}\relax
\EndOfBibitem
\bibitem[Tongay \latin{et~al.}(2013)Tongay, Suh, Ataca, Fan, Luce, Kang, Liu,
  Ko, Raghunathanan, Zhou, \latin{et~al.} others]{tongay2013defects}
Tongay,~S.; Suh,~J.; Ataca,~C.; Fan,~W.; Luce,~A.; Kang,~J.~S.; Liu,~J.;
  Ko,~C.; Raghunathanan,~R.; Zhou,~J.; others Defects activated
  photoluminescence in two-dimensional semiconductors: interplay between bound,
  charged and free excitons. \emph{Scientific reports} \textbf{2013}, \emph{3},
  2657\relax
\mciteBstWouldAddEndPuncttrue
\mciteSetBstMidEndSepPunct{\mcitedefaultmidpunct}
{\mcitedefaultendpunct}{\mcitedefaultseppunct}\relax
\EndOfBibitem
\bibitem[Srivastava \latin{et~al.}(2015)Srivastava, Sidler, Allain, Lembke,
  Kis, and Imamo{\u{g}}lu]{Srivastava2015}
Srivastava,~A.; Sidler,~M.; Allain,~A.~V.; Lembke,~D.~S.; Kis,~A.;
  Imamo{\u{g}}lu,~A. Optically active quantum dots in monolayer WSe2.
  \emph{Nature Nanotechnology} \textbf{2015}, \emph{10}, 491--496\relax
\mciteBstWouldAddEndPuncttrue
\mciteSetBstMidEndSepPunct{\mcitedefaultmidpunct}
{\mcitedefaultendpunct}{\mcitedefaultseppunct}\relax
\EndOfBibitem
\bibitem[Chen \latin{et~al.}(2023)Chen, Yue, Zhang, Xu, Liu, Feng, Hu, Yan,
  Scheuer, and Fu]{chen2023activated}
Chen,~X.; Yue,~X.; Zhang,~L.; Xu,~X.; Liu,~F.; Feng,~M.; Hu,~Z.; Yan,~Y.;
  Scheuer,~J.; Fu,~X. Activated Single Photon Emitters And Enhanced Deep-Level
  Emissions in Hexagonal Boron Nitride Strain Crystal. \emph{Advanced
  Functional Materials} \textbf{2023}, 2306128\relax
\mciteBstWouldAddEndPuncttrue
\mciteSetBstMidEndSepPunct{\mcitedefaultmidpunct}
{\mcitedefaultendpunct}{\mcitedefaultseppunct}\relax
\EndOfBibitem
\bibitem[Parto \latin{et~al.}(2021)Parto, Azzam, Banerjee, and
  Moody]{Parto2021}
Parto,~K.; Azzam,~S.~I.; Banerjee,~K.; Moody,~G. Defect and strain engineering
  of monolayer WSe2 enables site-controlled single-photon emission up to
  150{\thinspace}K. \emph{Nature Communications} \textbf{2021}, \emph{12},
  3585\relax
\mciteBstWouldAddEndPuncttrue
\mciteSetBstMidEndSepPunct{\mcitedefaultmidpunct}
{\mcitedefaultendpunct}{\mcitedefaultseppunct}\relax
\EndOfBibitem
\bibitem[Hern{\'a}ndez~L{\'o}pez \latin{et~al.}(2022)Hern{\'a}ndez~L{\'o}pez,
  Heeg, Schattauer, Kovalchuk, Kumar, Bock, Kirchhof, H{\"o}fer, Greben,
  Yagodkin, \latin{et~al.} others]{hernandez2022strain}
Hern{\'a}ndez~L{\'o}pez,~P.; Heeg,~S.; Schattauer,~C.; Kovalchuk,~S.;
  Kumar,~A.; Bock,~D.~J.; Kirchhof,~J.~N.; H{\"o}fer,~B.; Greben,~K.;
  Yagodkin,~D.; others Strain control of hybridization between dark and
  localized excitons in a 2D semiconductor. \emph{Nature communications}
  \textbf{2022}, \emph{13}, 7691\relax
\mciteBstWouldAddEndPuncttrue
\mciteSetBstMidEndSepPunct{\mcitedefaultmidpunct}
{\mcitedefaultendpunct}{\mcitedefaultseppunct}\relax
\EndOfBibitem
\bibitem[Vutukuru \latin{et~al.}(2021)Vutukuru, Ardekani, Chen, Wilmington,
  Gundogdu, and Swan]{vutukuru2021enhanced}
Vutukuru,~M.; Ardekani,~H.; Chen,~Z.; Wilmington,~R.~L.; Gundogdu,~K.;
  Swan,~A.~K. Enhanced dielectric screening and photoluminescence from
  nanopillar-strained MoS2 nanosheets: Implications for strain funneling in
  optoelectronic applications. \emph{ACS Applied Nano Materials} \textbf{2021},
  \emph{4}, 8101--8107\relax
\mciteBstWouldAddEndPuncttrue
\mciteSetBstMidEndSepPunct{\mcitedefaultmidpunct}
{\mcitedefaultendpunct}{\mcitedefaultseppunct}\relax
\EndOfBibitem
\bibitem[Harats \latin{et~al.}(2020)Harats, Kirchhof, Qiao, Greben, and
  Bolotin]{harats2020}
Harats,~M.~G.; Kirchhof,~J.~N.; Qiao,~M.; Greben,~K.; Bolotin,~K.~I. Dynamics
  and efficient conversion of excitons to trions in non-uniformly strained
  monolayer WS2. \emph{Nature Photonics} \textbf{2020}, \emph{14},
  324--329\relax
\mciteBstWouldAddEndPuncttrue
\mciteSetBstMidEndSepPunct{\mcitedefaultmidpunct}
{\mcitedefaultendpunct}{\mcitedefaultseppunct}\relax
\EndOfBibitem
\bibitem[Iff \latin{et~al.}(2019)Iff, Tedeschi, Mart{\'i}n-S{\'a}nchez,
  Mocza{\l}a-Dusanowska, Tongay, Yumigeta, Taboada-Guti{\'e}rrez, Savaresi,
  Rastelli, Alonso-Gonz{\'a}lez, H{\"o}fling, Trotta, and Schneider]{Iff2019}
Iff,~O.; Tedeschi,~D.; Mart{\'i}n-S{\'a}nchez,~J.; Mocza{\l}a-Dusanowska,~M.;
  Tongay,~S.; Yumigeta,~K.; Taboada-Guti{\'e}rrez,~J.; Savaresi,~M.;
  Rastelli,~A.; Alonso-Gonz{\'a}lez,~P.; H{\"o}fling,~S.; Trotta,~R.;
  Schneider,~C. Strain-Tunable Single Photon Sources in WSe2 Monolayers.
  \emph{Nano Letters} \textbf{2019}, \emph{19}, 6931--6936\relax
\mciteBstWouldAddEndPuncttrue
\mciteSetBstMidEndSepPunct{\mcitedefaultmidpunct}
{\mcitedefaultendpunct}{\mcitedefaultseppunct}\relax
\EndOfBibitem
\bibitem[Kirchhof \latin{et~al.}(2022)Kirchhof, Yu, Antheaume, Gordeev,
  Yagodkin, Elliott, De~Ara{\'u}jo, Sharma, Reich, and
  Bolotin]{kirchhof2022nanomechanical}
Kirchhof,~J.~N.; Yu,~Y.; Antheaume,~G.; Gordeev,~G.; Yagodkin,~D.; Elliott,~P.;
  De~Ara{\'u}jo,~D.~B.; Sharma,~S.; Reich,~S.; Bolotin,~K.~I. Nanomechanical
  spectroscopy of 2d materials. \emph{Nano letters} \textbf{2022}, \emph{22},
  8037--8044\relax
\mciteBstWouldAddEndPuncttrue
\mciteSetBstMidEndSepPunct{\mcitedefaultmidpunct}
{\mcitedefaultendpunct}{\mcitedefaultseppunct}\relax
\EndOfBibitem
\bibitem[Warburton \latin{et~al.}(2000)Warburton, Sch{\"a}flein, Haft, Bickel,
  Lorke, Karrai, Garcia, Schoenfeld, and Petroff]{warburton2000optical}
Warburton,~R.~J.; Sch{\"a}flein,~C.; Haft,~D.; Bickel,~F.; Lorke,~A.;
  Karrai,~K.; Garcia,~J.~M.; Schoenfeld,~W.; Petroff,~P.~M. Optical emission
  from a charge-tunable quantum ring. \emph{Nature} \textbf{2000}, \emph{405},
  926--929\relax
\mciteBstWouldAddEndPuncttrue
\mciteSetBstMidEndSepPunct{\mcitedefaultmidpunct}
{\mcitedefaultendpunct}{\mcitedefaultseppunct}\relax
\EndOfBibitem
\bibitem[Tran \latin{et~al.}(2016)Tran, Bray, Ford, Toth, and
  Aharonovich]{tran2016quantum}
Tran,~T.~T.; Bray,~K.; Ford,~M.~J.; Toth,~M.; Aharonovich,~I. Quantum emission
  from hexagonal boron nitride monolayers. \emph{Nature nanotechnology}
  \textbf{2016}, \emph{11}, 37--41\relax
\mciteBstWouldAddEndPuncttrue
\mciteSetBstMidEndSepPunct{\mcitedefaultmidpunct}
{\mcitedefaultendpunct}{\mcitedefaultseppunct}\relax
\EndOfBibitem
\bibitem[He \latin{et~al.}(2016)He, Iff, Lundt, Baumann, Davanco, Srinivasan,
  H{\"o}fling, and Schneider]{He2016}
He,~Y.-M.; Iff,~O.; Lundt,~N.; Baumann,~V.; Davanco,~M.; Srinivasan,~K.;
  H{\"o}fling,~S.; Schneider,~C. Cascaded emission of single photons from the
  biexciton in monolayered WSe2. \emph{Nature Communications} \textbf{2016},
  \emph{7}, 13409\relax
\mciteBstWouldAddEndPuncttrue
\mciteSetBstMidEndSepPunct{\mcitedefaultmidpunct}
{\mcitedefaultendpunct}{\mcitedefaultseppunct}\relax
\EndOfBibitem
\bibitem[Darlington \latin{et~al.}(2020)Darlington, Carmesin, Florian, Yanev,
  Ajayi, Ardelean, Rhodes, Ghiotto, Krayev, Watanabe, \latin{et~al.}
  others]{darlington2020}
Darlington,~T.~P.; Carmesin,~C.; Florian,~M.; Yanev,~E.; Ajayi,~O.;
  Ardelean,~J.; Rhodes,~D.~A.; Ghiotto,~A.; Krayev,~A.; Watanabe,~K.; others
  Imaging strain-localized excitons in nanoscale bubbles of monolayer WSe2 at
  room temperature. \emph{Nature Nanotechnology} \textbf{2020}, \emph{15},
  854--860\relax
\mciteBstWouldAddEndPuncttrue
\mciteSetBstMidEndSepPunct{\mcitedefaultmidpunct}
{\mcitedefaultendpunct}{\mcitedefaultseppunct}\relax
\EndOfBibitem
\bibitem[Kim \latin{et~al.}(2019)Kim, Moon, Noh, Lee, and Kim]{kim2019}
Kim,~H.; Moon,~J.~S.; Noh,~G.; Lee,~J.; Kim,~J.-H. Position and frequency
  control of strain-induced quantum emitters in WSe2 monolayers. \emph{Nano
  letters} \textbf{2019}, \emph{19}, 7534--7539\relax
\mciteBstWouldAddEndPuncttrue
\mciteSetBstMidEndSepPunct{\mcitedefaultmidpunct}
{\mcitedefaultendpunct}{\mcitedefaultseppunct}\relax
\EndOfBibitem
\bibitem[Aslan \latin{et~al.}(2018)Aslan, Deng, and Heinz]{aslan2018strain}
Aslan,~B.; Deng,~M.; Heinz,~T.~F. Strain tuning of excitons in monolayer WSe 2.
  \emph{Physical Review B} \textbf{2018}, \emph{98}, 115308\relax
\mciteBstWouldAddEndPuncttrue
\mciteSetBstMidEndSepPunct{\mcitedefaultmidpunct}
{\mcitedefaultendpunct}{\mcitedefaultseppunct}\relax
\EndOfBibitem
\bibitem[Moon \latin{et~al.}(2020)Moon, Bersin, Chakraborty, Lu, Grosso, Kong,
  and Englund]{moon2020strain}
Moon,~H.; Bersin,~E.; Chakraborty,~C.; Lu,~A.-Y.; Grosso,~G.; Kong,~J.;
  Englund,~D. Strain-correlated localized exciton energy in atomically thin
  semiconductors. \emph{ACS Photonics} \textbf{2020}, \emph{7},
  1135--1140\relax
\mciteBstWouldAddEndPuncttrue
\mciteSetBstMidEndSepPunct{\mcitedefaultmidpunct}
{\mcitedefaultendpunct}{\mcitedefaultseppunct}\relax
\EndOfBibitem
\bibitem[Conley and et~al.(2015)Conley, and et~al.]{nr2015}
Conley,~H.~J.; et~al. Strain Effects on the Electronic Properties of Monolayer
  Transition Metal Dichalcogenides. \emph{Nano Research} \textbf{2015},
  \emph{8}, 2562--2572\relax
\mciteBstWouldAddEndPuncttrue
\mciteSetBstMidEndSepPunct{\mcitedefaultmidpunct}
{\mcitedefaultendpunct}{\mcitedefaultseppunct}\relax
\EndOfBibitem
\bibitem[Korn and et~al.(2013)Korn, and et~al.]{nl2013}
Korn,~T.; et~al. Direct-to-Indirect Bandgap Transition in Monolayer Transition
  Metal Dichalcogenides. \emph{Nano Letters} \textbf{2013}, \emph{13},
  4095--4101\relax
\mciteBstWouldAddEndPuncttrue
\mciteSetBstMidEndSepPunct{\mcitedefaultmidpunct}
{\mcitedefaultendpunct}{\mcitedefaultseppunct}\relax
\EndOfBibitem
\bibitem[Fei and Yang(2017)Fei, and Yang]{npj2D2017}
Fei,~R.; Yang,~L. Strain Engineering in Two-Dimensional Semiconductors.
  \emph{npj 2D Materials and Applications} \textbf{2017}, \emph{1}, 10\relax
\mciteBstWouldAddEndPuncttrue
\mciteSetBstMidEndSepPunct{\mcitedefaultmidpunct}
{\mcitedefaultendpunct}{\mcitedefaultseppunct}\relax
\EndOfBibitem
\bibitem[Wang and et~al.(2022)Wang, and et~al.]{aip2022}
Wang,~S.; et~al. Bandgap Modulation in Transition Metal Dichalcogenides under
  Tensile Strain. \emph{AIP Advances} \textbf{2022}, \emph{12}, 115023\relax
\mciteBstWouldAddEndPuncttrue
\mciteSetBstMidEndSepPunct{\mcitedefaultmidpunct}
{\mcitedefaultendpunct}{\mcitedefaultseppunct}\relax
\EndOfBibitem
\bibitem[He and et~al.(2014)He, and et~al.]{rsc2014}
He,~J.; et~al. Electronic and Optical Properties of Transition Metal
  Dichalcogenides under Strain. \emph{RSC Advances} \textbf{2014}, \emph{4},
  34561--34567\relax
\mciteBstWouldAddEndPuncttrue
\mciteSetBstMidEndSepPunct{\mcitedefaultmidpunct}
{\mcitedefaultendpunct}{\mcitedefaultseppunct}\relax
\EndOfBibitem
\end{mcitethebibliography}

\newpage
\begin{suppinfo}

%%%%%%%%%%%%%%%%%%%%%%%%%%%%%%%%%%%%%%%%%%%%%%%%%%%%%%%%%%%%%%%%%%%
\renewcommand{\figurename}{Supplementary Figure}
\setcounter{figure}{0}
\begin{figure}
	\centering
	\includegraphics[width=1\textwidth]{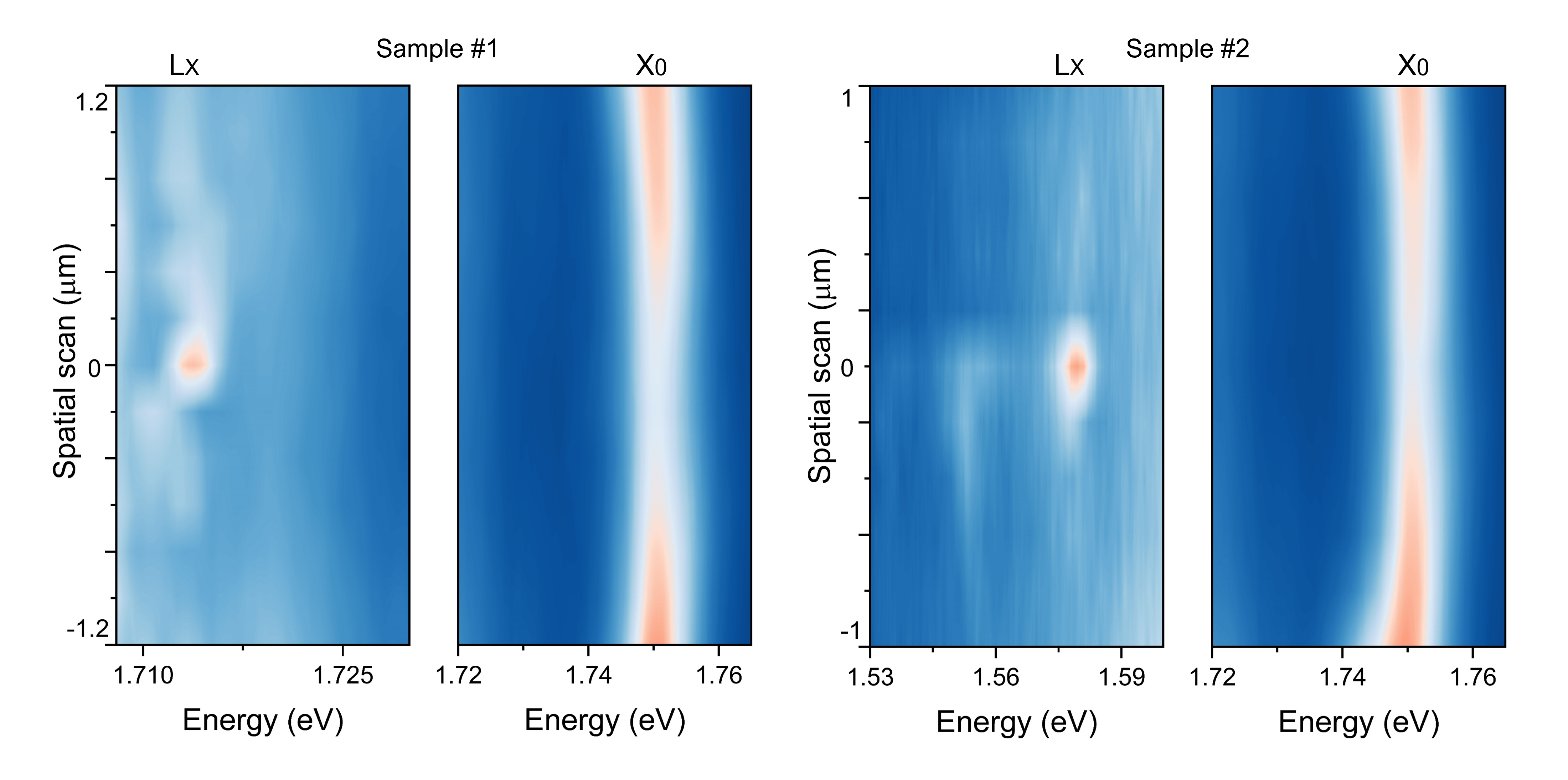}
	\caption{ \textbf{Localized character of the emission.} PL intensity vs. spatial position across the indentation region. The slight shift in the neutral exciton spectral position is due to strain variation across the flake. The maximum indentation force is 6 $\mu$N in sample-1 and sample-2 (shown on top of the stack spectra in Fig. 3a).}
\end{figure}
%------------------------------------------------------------

\renewcommand{\figurename}{Supplementary Figure}

%---------------------------------------------

\renewcommand{\figurename}{Supplementary Figure}
\begin{figure}
	\centering
	\includegraphics[width=1\textwidth]{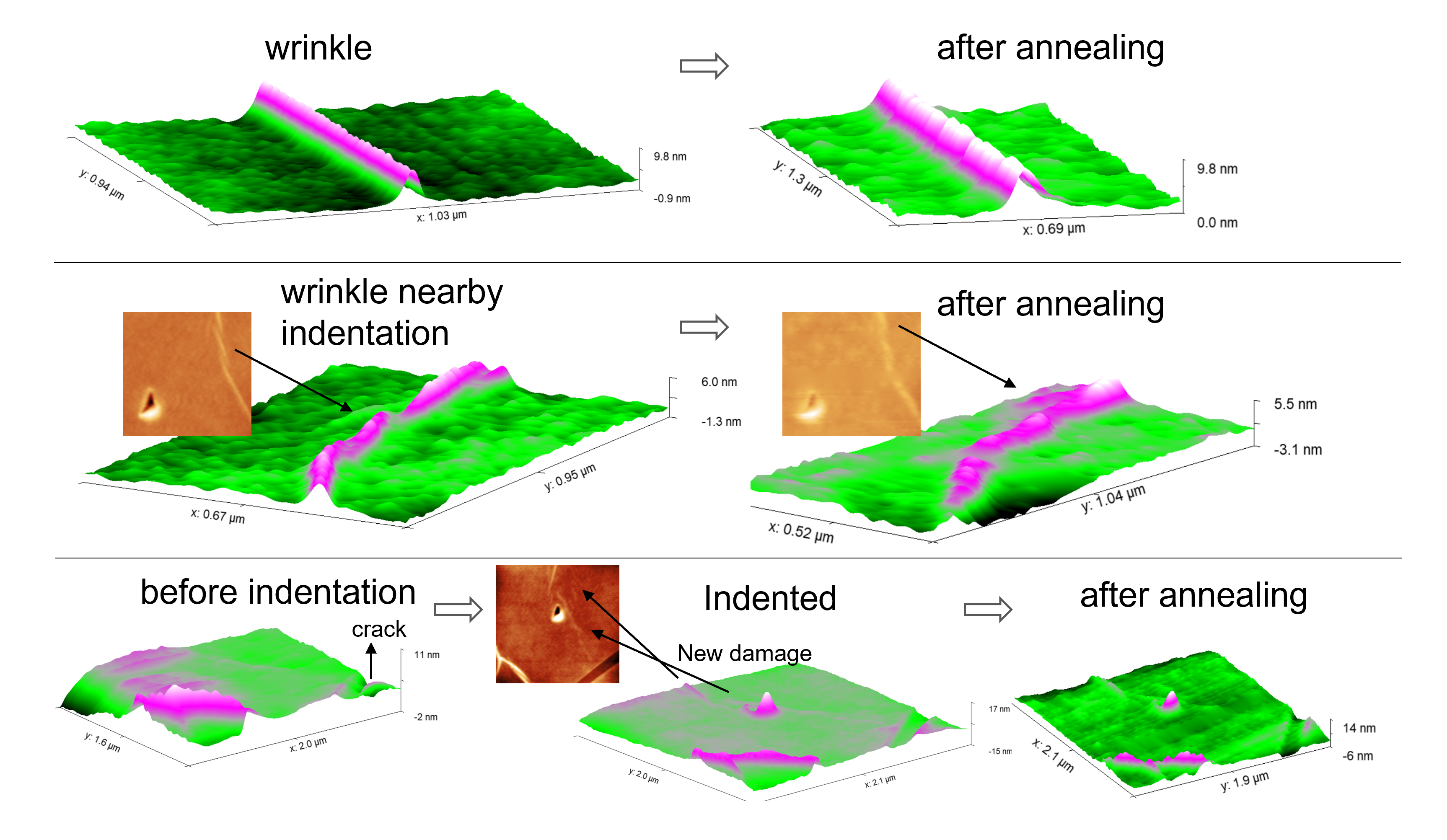}
	\caption{ Examples of irregularities on the \ce{WSe2} flake }
\end{figure}
%-----------------------------------------------------------
%------------------------------------------------------------
\renewcommand{\figurename}{Supplementary Figure}
\begin{figure}
	\centering
	\includegraphics[width=1\textwidth]{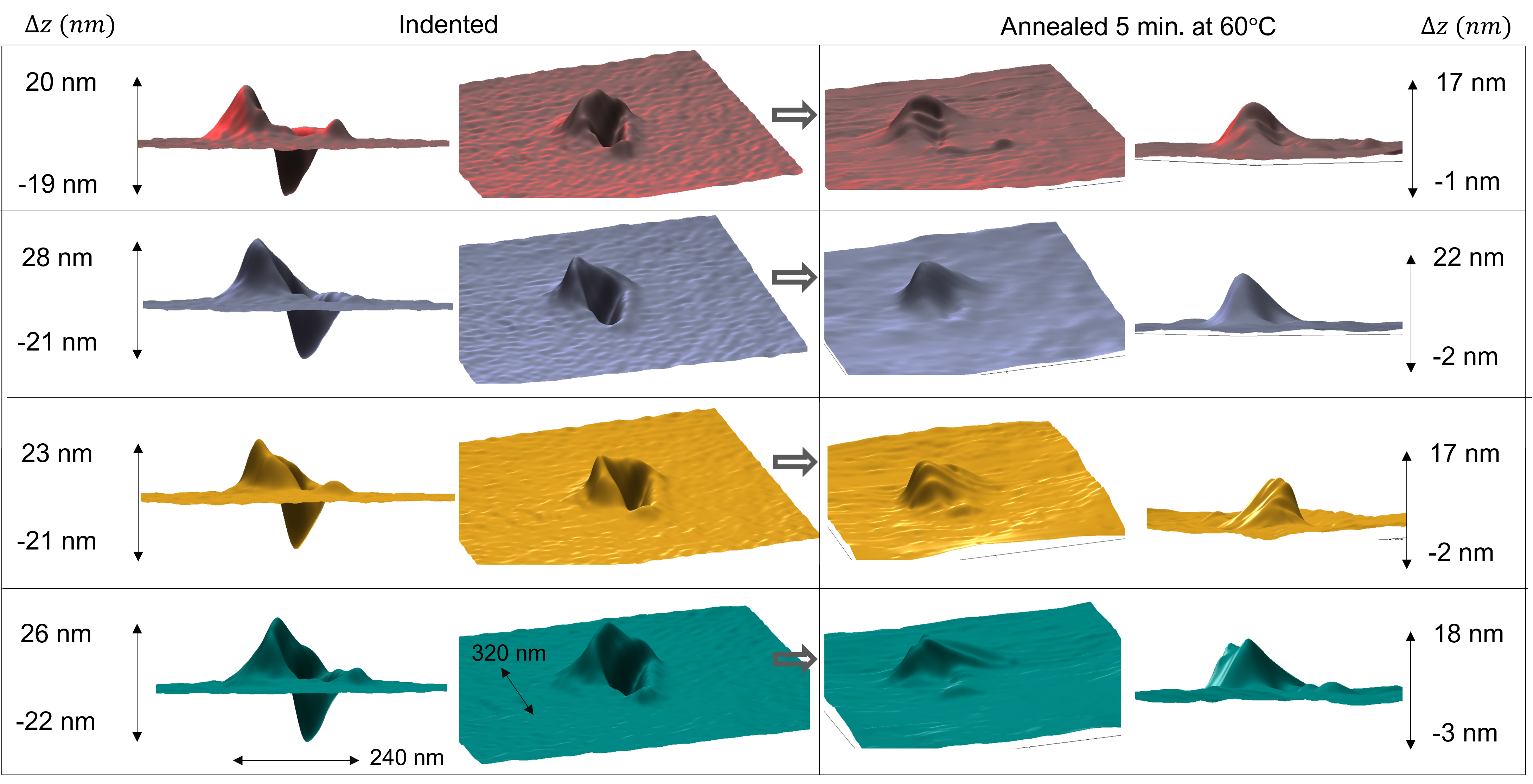}
	\caption{Additional high-resolution AFM image examples of indentation profiles before and after annealing. }
\end{figure}
%-----------------------------------------------------------
%------------------------------------------------------------
\renewcommand{\figurename}{Supplementary Figure}
\begin{figure}
	\centering
	\includegraphics[width=1\textwidth]{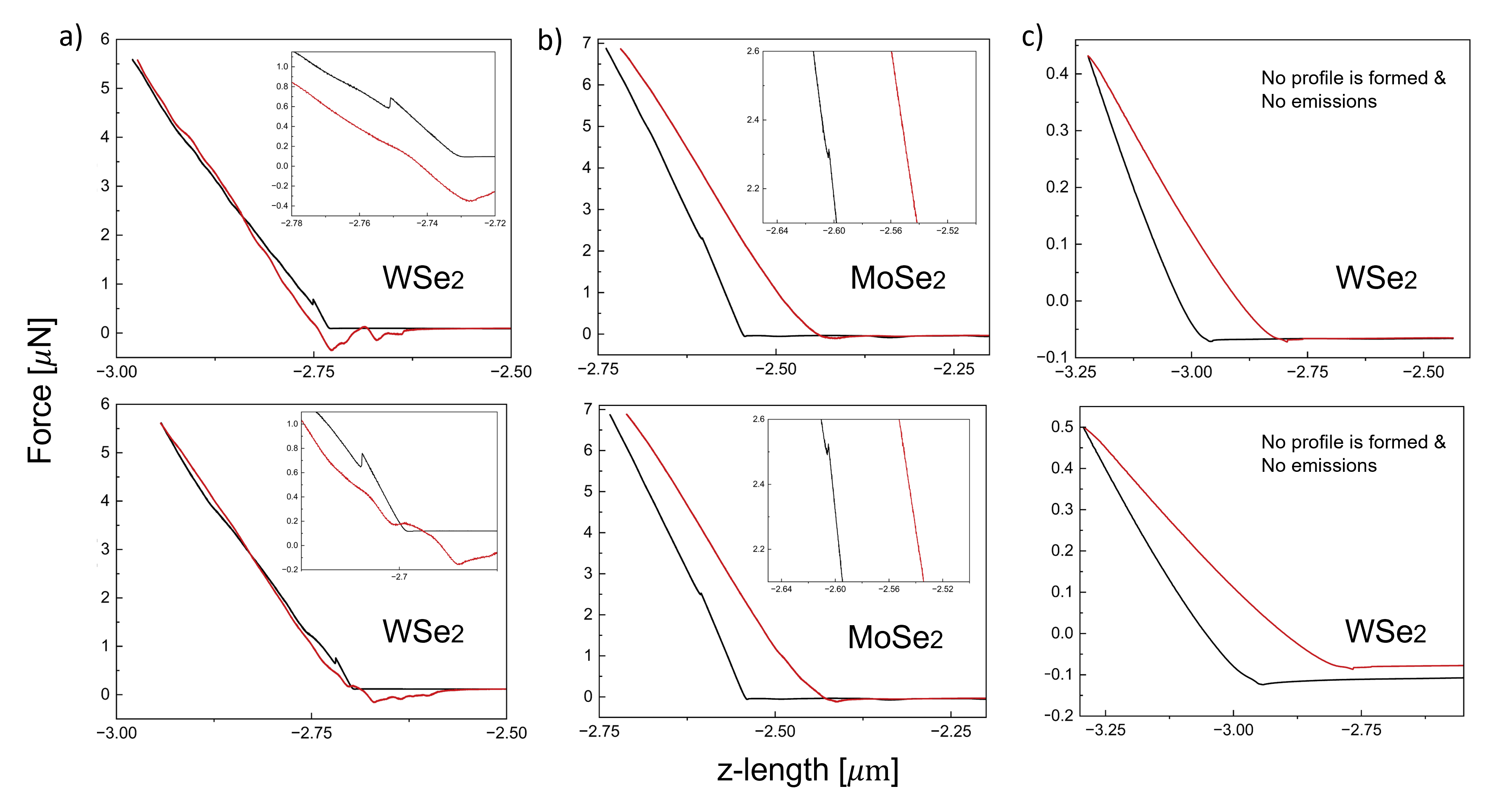}
	\caption{Additional examples of indentation force curves for different samples. Black lines are extension and red curves are retraction.}
\end{figure}
%----------------------------------------------
%------------------------------------------------------------
\renewcommand{\figurename}{Supplementary Figure}
\begin{figure}
	\centering
	\includegraphics[width=1\textwidth]{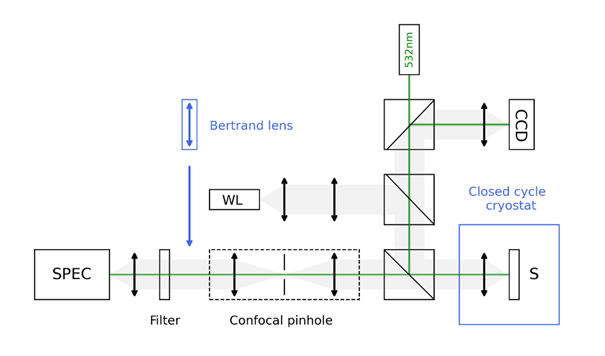}
	\caption{The schematics of the measurement setup.}
\end{figure}
%-----------------------------------------------

\end{suppinfo}


\end{document}